\documentclass[a4paper]{article}

\usepackage{ISCSLP2024}
\usepackage{float} % 添加这个包来使用 H 位置参数
\usepackage{tipa} % 添加对国际音标的支持

\usepackage{CJKutf8} % 添加对中文的支持
\usepackage{makecell}

\title{Low-Resourced Speech Recognition for Iu Mien Language via  Weakly-Supervised Phoneme-based Multilingual Pre-training}
\name{Lukuan Dong$^1$, Donghong Qin$^1$, Fengbo Bai$^1$, Fanhua Song$^1$,Yan Liu$^2$, Chen Xu$^1$, Zhijian Ou$^2$ \thanks{
Corresponding author and project lead: D. Qin and Z. Ou. 
 This work is supported by Guangxi Science and Technology Base and Talent Project (GUIKEAD23026054).
}
}

\address{
  $^1$AI and Big Data International Cooperation Joint Laboratory, School of Artificial Intelligence, 
Guangxi Minzu University, Nanning, China \\
  $^2$Speech Processing and Machine Intelligence (SPMI) Lab, Tsinghua University, Beijing, China}
\email{donglukuan@stu.gxmzu.edu.cn, ozj@tsinghua.edu.cn}

\begin{document}

\begin{CJK*}{UTF8}{gbsn} % 中文支持

\renewcommand{\arraystretch}{1.1} % 调整表格行高

\maketitle
\begin{abstract}
The mainstream automatic speech recognition (ASR) technology usually requires hundreds to thousands of hours of annotated speech data. Three approaches to low-resourced ASR are phoneme or subword based supervised pre-training, and self-supervised pre-training over multilingual data. The Iu Mien language is the main ethnic language of the Yao ethnic group in China and is low-resourced in the sense that the annotated speech is very limited. With less than 10 hours of transcribed Iu Mien language, this paper investigates and compares the three approaches for Iu Mien speech recognition. Our experiments are based on the recently released, three backbone models pretrained over the 10 languages from the CommonVoice dataset (CV-Lang10), which correspond to the three approaches for low-resourced ASR. It is found that phoneme supervision can achieve better results compared to subword supervision and self-supervision, thereby providing higher data-efficiency. Particularly, the Whistle models, i.e., obtained by the weakly-supervised phoneme-based multilingual pre-training, obtain the most competitive results.
\end{abstract}

\section{Introduction}
The Yao ethnic group is one of the ethnic minorities in China, with a population of over 3 million people, mainly distributed in Guangxi, Guangdong, Hunan, Guizhou, Yunnan and other regions. It is the most widely distributed ethnic minority in southern China, and also exists in Southeast Asian countries such as Vietnam, Thailand, and Cambodia. Yao is the language of the Yao ethnic group, but it lacks a completely unified standardization and consists of many dialects. Among these, Iu Mien is a major dialect and serves as the basis for the current official standard Yao language. Iu Mien belongs to the Yao branch of the Iu Mien-Yao sub-family within the Sino-Tibetan language family. Its written form is a phonetic syllabic system using basic Latin letters.

With the development of deep learning based artificial intelligence, the capabilities of automatic speech recognition (ASR) technology have also developed rapidly. However, current speech recognition technology often requires hundreds to thousands of hours of training data to obtain good performances. Due to the long-term scattered residence of the Yao population and the lack of an authoritative central dialect, the speech and language resources of the Iu Mien language are very scarce. In our practice, it takes non-trivial efforts to collect and transcribe even less than 10 hours of Iu Mien language. The development of Iu Mien language speech recognition systems is very challenging, while it is very important to reduce digital divides and culture inheritance. 

The paradigm of pre-training (PT) followed by fine-tuning (FT), called the PTFT paradigm, has emerged in recent years as an effective way to solve the problem of limited training data for low-resource languages for ASR. 
%【说完PT FT emerged，要解释，娓娓道来】
In pre-training, training data for a number of languages are merged to train a multilingual model. The pre-trained model can then serve as a backbone, which can be further fine-tuned for crosslingual speech recognition. This PTFT paradigm helps the backbone to learn common knowledge for speech recognition from multiple, different languages during pre-training. So when fine-tuning the backbone model on the target language, decent ASR performance could be obtained with only a small amount of target language training data. 

%以下重写：
%引出three approaches及围绕“本文立意简要评述”（更多介绍放在Related work）。
%介绍IPA的好处，介绍whistle，然后介绍本文工作及贡献

Currently, there are three main multilingual pre-training methods: self-supervised pre-training, subword-based supervised pre-training, and phoneme-based supervised pre-training. Self-supervised pre-training uses unlabeled audio data to train models to learn common representations of multilingual audio. Subword-based pre-training combines multilingual annotated audio data by creating a shared token set for multiple languages. Phoneme-based supervised pre-training typically uses International Phonetic Alphabet (IPA) as a common pronunciation annotation system, enabling effective multilingual supervised pre-training. The IPA is designed to provide a unified system of symbols to represent the basic sounds of different languages. Each IPA symbol corresponds to a specific phoneme, ensuring a one-to-one relationship between the symbol and the sound. With the IPA, it is possible to achieve consistent phonetic transcription across all languages \cite{fromkin2007introduction}.
Comparing the three pre-training methods, the phoneme-based approach not only allows for efficient model training but also maximizes the sharing of pronunciation features across different languages. However, obtaining high-quality IPA pronunciation annotations for many minor languages is challenging.

Our work is based on a recent study on weakly supervised phoneme pretraining, Whistle \cite{yusuyin2024whistle}. 
In \cite{yusuyin2024whistle}, the effects of different pre-training methods on multilingual and cross-language speech recognition are analyzed and compared. A weakly phonetic supervision method (Whistle) is proposed. It is shown in \cite{yusuyin2024whistle} that phoneme-based supervised pre-training models are more data-efficient, revealing that they perform better in cross-language speech recognition with limited training data. Additionally, related pre-training models are publicly released, which we utilize to construct a Iu Mien language speech recognition model in this work. 

To the best of our knowledge, there has been no studies on using the pre-training and fine-tuning paradigm to train Iu Mien language speech recognition models. The aim of this paper is to explore the potential of the Whistle modeling on Iu Mien Language, a Sino-Tibetan low-resource language. Specifically, this paper uses the Whistle model as a pre-trained backbone model and fine-tuning it with a small amount of Iu Mien language training data to construct a Iu Mien language speech recognition model. We compare the fine-tuning results on Iu Mien with those obtained using backbone models from other pre-training methods. Our main contributions are as follows.

\begin{itemize}
\item This paper introduces the basic characteristics of the Iu Mien language, including its writing and pronunciation features, and explores how to train a Iu Mien language speech recognition model using a small amount of data.
\item Experiments demonstrate that the Whistle method is more data-efficient compared to other methods, resulting in better Iu Mien language speech recognition models with an limited amount of Iu Mien speech data.

\end{itemize}

\begin{table*}[th]
    \caption{Examples of Iu Mien language word spellings.}
    \label{tab:spelling}
    \centering
    \begin{tabular}{ccccccccc}
        \toprule
        \textbf{Iu Mien word} & \textbf{Chinese meaning} & \textbf{Chinese meaning} & \textbf{syllable} & \textbf{initial consonant} & \multicolumn{3}{c}{\textbf{vowel}}  &  \textbf{tone} \\
        \cmidrule(lr){6-8}
        & & \textbf{in English} & & & \textbf{initial} & \textbf{main} & \textbf{final} & \\
        \midrule
        ginghgungv & \makecell[c]{蜻蜓}
 & dragonfly & gingh & g & - & i & ng & h \\
        \hline
         & & & gungv & g & - & u & ng & v \\
         \hline
        baengh & \makecell[c]{平}
 & flat & baengh & b & - & ae & ng & h \\
        \hline
        nqaang & \makecell[c]{后来}
 & later & nqaang & nq & - & aa & ng & - \\
        \hline
        guinh & \makecell[c]{圈，转}
 & circle, turn & guinh & g & u & i & n & h \\
        \bottomrule
    \end{tabular}
\end{table*}

\section{Related work}
Grapheme-based pre-training methods use graphemes as modeling units to construct a shared vocabulary for multilingual data. There are currently three basic grapheme modeling methods: characters \cite{8682256}, subwords \cite{zheng2021advancing}, and words \cite{soltau17_interspeech}. To retain some semantic information while avoiding the out-of-vocabulary (OOV) issue, the most commonly used method is to use subwords for modeling. For example, Whisper \cite{10.5555/3618408.3619590} uses BPE-based (Byte-Pair Encoding base) text tokenizer and weakly graphemic supervision on more than 680,000 hours of clean web data, which can recognize more than 97 languages. However, graphemes are related to the writing system of a language, and it may be difficult to train grapheme based models to learn common speech recognition capabilities between different languages. Self-supervised training is another major multilingual pre-training method. For example, models such as XLS-R \cite{Babu2021XLSRSC} use a large amount of unlabeled multilingual audio data for training, and their training methods are similar to wav2vec 2.0 \cite{baevski2020wav2vec} or BERT \cite{kenton2019bert}. Based on the pre-trained model, a linear layer is added to map the output to the target token list of speech recognition. The CTC approach  \cite{graves2006connectionist} can then be used to fine-tune the model.

Supervised pre-training based on phoneme annotations is different from supervised pre-training methods based on graphemics, in that it can reduce the influence of the writing systems of different languages during model training. On the other hand, compared with self-supervised training, clear phoneme training goals can also help the model quickly learn general speech recognition knowledge. However, obtaining accurate pronunciation annotations for each language often requires a lot of expert knowledge. Recently,  \cite{yusuyin2024whistle} proposed a new phoneme-based supervised pre-training method called Whistle. Whistle uses IPA phoneme annotations with a small number of errors generated by the G2P model as the target (weakly supervision) during model pre-training to train a multilingual speech recognition model. Whistle used data from ten languages (English, French, Italian, Spanish, Russian, Dutch, Turkish, Kyrgyz, Swedish, Tatar) from commonvoice11 \cite{ardila2020common} as training data. 
% The G2P model is used to generate corresponding IPA annotations based on text annotations and removed the diacritics from the IPA annotations. 
The Conformer model was used as the encoder, and the CTC algorithm was used for supervised pre-training. Finally, three pre-trained models of different sizes were trained: Whistle-small (90M), Whistle-medium (220M), and Whistle-large (543M).  \cite{yusuyin2024whistle} conducted cross-language fine-tuning experiments on Polish and Indonesian, respectively, to demonstrate the effectiveness of this method. However, during the pre-training process of  \cite{yusuyin2024whistle}, no Sino-Tibetan language was incorporated as the pre-training languages, and the IPA annotations of the pre-training data did not contain tone information.

\section{Iu Mien language spelling scheme}

The writing system currently used by the Yao people is a newly created writing system, which was unified by China and the United States in 1984. It spells the Iu Mien dialect. Therefore, it can also be called the Iu Mien Unified Script (IMUC).
The script is a phonetic syllabic script based on the Latin alphabet, and each syllable can be composed of 30 initials, 128 finals, and 8 tones. As shown in Table \ref{tab:spelling}, IMUC is a phonetic script that uses subwords to represent different sounds. For each IMUC word, its structure is similar to the Chinese pinyin to represent words, consisting of one or more syllables, but unlike Chinese pinyin, Iu Mien only uses the 26 letters of the basic Latin alphabet, and does not use additional symbols to distinguish tones or vowels. IMUC generally uses Latin letters at the end of the word to represent tones. However, there is no corresponding letter for the mid-level tone. If a word ends without a tone-indicating letter, it indicates that the word has a mid-level tone. Other tones are represented by five letters: 'h', 'v', 'z', 'x', and 'c'.

\begin{figure}[t]
    \centering
    \includegraphics[width=\linewidth]{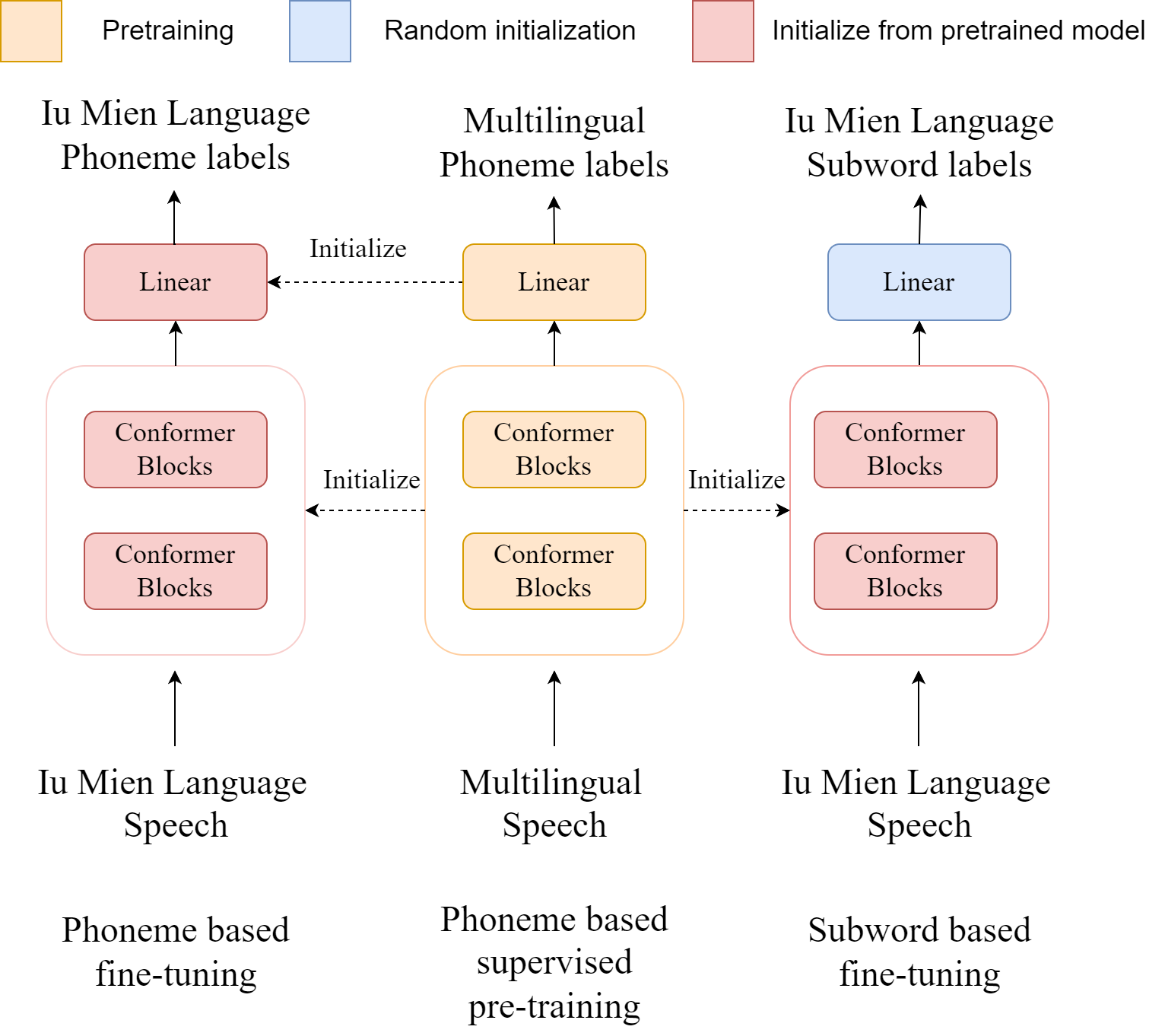}
    \caption{Illustration of the fine-tuning procedures with phonetic supervision pre-training.}
    \label{fig:methods}
\end{figure}

\section{Method}

This paper use a weakly-supervised phoneme-based multilingual pre-training model and then fine-tuning to improve the recognition performance of the Iu Mien language speech recognition model. As shown in Figure \ref{fig:methods}, we use an end-to-end speech recognition model based on the CTC approach. The acoustic encoder of the Iu Mien language speech recognition model is initialized using the pre-trained Whistle-small model. The fine-tuning training of the Iu Mien language speech recognition model is performed separately based on subword modeling and phoneme modeling.

We add a linear layer on top of the acoustic encoder, which is used to map the feature vectors of the encoder output to the target prediction space. For methods of fine-tuning based on subwords, we use the random initialization method to initialize the linear layer parameters. For the phoneme-based approach to fine-tuning, in order to share the parameters from the pre-training as much as possible, we used the method in Whistle to use the parameters of the linear layer from the pre-training backbone for initialization. For the k-th row vector in the linear layer weight matrix at the top of the encoder during pre-training, it is considered as the embedding vector corresponding to the k-th phoneme in the predicted phoneme list \cite{yusuyin2024whistle}. For a target language phoneme that has appeared in the predicted phoneme lexicon at pre-training, we use the embedding vector of that phoneme obtained at pre-training to initialize the corresponding position of the linear layer weight matrix at fine-tuning.

\section{Experimental setup}

\subsection{Dataset}
MightLJSpeech is a publicly available dataset of the Iu Mien language. The dataset is 9.7 hours long, consisting of 9761 utterance from a female speaker and the corresponding Iu Mien language text annotations. The sampling rate for audio files is 22.05kHz. 

In the experiment, all audio files were first resampled to 16KHz, and then the dataset was divided into training set, development set and test set in a ratio of 8:1:1. To test the statistical significance of our experimental results, we conducted each experiment three independent times and used the average of these independent runs as the final result. For each run, we divided the dataset using a cross-validation approach. Specifically, we split the original dataset into 10 parts. In each run, two parts were used as the development set and test set, while the remaining eight parts were used as the training set. The development set and test set selected for each run were not reused in other runs for development or test.

\subsection{Pronunciation lexicon construction}
The file that records words and their corresponding IPA pronunciation annotations is called a pronunciation lexicon. For speech recognition models that use phoneme modeling, we need to incorporate the pronunciation dictionary to build a phoneme vocabulary. In phoneme-based speech recognition models, decoding often requires the use of a lexicon and a language model. In the experiment, the training set is used to generate a word vocabulary. Based on the IMUS-IPA spelling correspondence table provided in the Iu Mien Language Wikipedia, the corresponding IPA pronunciation annotation is generated for each word according to the longest match principle. For the tone of each word, numbers are used in the pronunciation lexicon to represent different tones.
\subsection{Model training}
In our experiment, the CAT toolkit  \cite{An2020CATAC} was used to train a CTC-based speech recognition model. The same model structure as in Whistle was used to facilitate the transfer of pre-trained model parameters and fine-tuning on the Iu Mien language training set. The acoustic encoder used a Conformer network model with 14 encoder blocks, each self-attention layer containing 4 self-attention heads, each with 36-dimensional hidden states. In the Whistle model based experiment, 80 fbank features extracted from the audio (16KHz resample) were used as audio input.

For the Iu Mien Language speech recognition experiments based on subword modeling, we use the BPE algorithm implemented through the SentencePiece \cite{kudo2018sentencepiece} tool to construct the tokenizer and set the vocabulary size to 500. For the Iu Mien Language speech recognition experiments based on phoneme modeling, the size of the phoneme vocabulary size is 54, and the size of the phoneme vocabulary after removing diacritics other than tones is 44.

In the experiment, early stopping was used as the training scheduling strategy. When the loss of the model on the development set did not decrease for a certain number of consecutive times, the training was stopped. Meanwhile, in order to make a fair comparison, we set a minimum number of training iterations so that different models are fully trained when conducting experimental comparisons.

\subsection{Model decoding}

Beam search with a beam size of 32 was used for decoding. In order to use the language model to improve the decoding with the speech recognition models, we used the Kenlm tool \cite{heafield2011kenlm} to train a word-level 4-gram language model based on the text of the Iu Mien Language training set. In the experiments based on subword modeling, we use WFST (weighted finite state transducer) based decoding \cite{mohri2008speech,8682256} to combine the acoustic model as well as the language model to improve the speech recognition results. In the experiments based on phoneme modeling, we use WFST-based decoding to combine the acoustic model, the pronunciation lexicon, and the language model to get the final decoding results.

\section{Experimental results}

\subsection{Results for subword-based FT}

Table \ref{tab:subword-result} shows the results for subword-based FT, starting from different pre-trained models.
O1 denotes the subword-based monolingual baseline, i.e., trained from scratch.
M1 denotes the result from fine-tuning of Whistle-small.
M3 denotes the result from fine-tuning of a Wav2Vec2-base model, trained from scratch using self-supervised pre-training with the same training data as the Whistle-small model, which we call the Wav2Vec2-cv10 model.
% As shown in Table \ref{tab:subword-result}, we compare M1 with a Wav2Vec2-base model trained from scratch using self-supervised pre-training with the same training data as the Whistle-small model, which we call the Wav2Vec2-cv10 model. 
Wav2Vec2-cv10 was trained using the fairseq toolkit, following the wav2vec 2.0 base pre-training configuration provided with the toolkit. 
M4 denotes the result from fine-tuning of a subword pretraining model, called Mul10-subword, which uses the same pretraining data and encoder architecture as Whistle-small, but is modeled using subwords with a vocabulary size of 5000. To fairly compare the performance of models trained with the same pretraining data using different methods, the parameter count of Wav2Vec2-cv10 and Mul10-subword is kept similar to Whistle-small (90M).
To show the statistical significance of the experimental results, we conducted three independent cross-validation tests for each experiment and took the average as the final experimental result.

\begin{table}[t]
  \caption{WERs for subword-based fine-tuning (averaged over three independent cross-validation runs). O and M are shorthands for m\underline{o}nolingual and \underline{m}ultilingual models.}
  \label{tab:subword-result}
  \centering
  \begin{tabular}{ cp{2cm}cc }
    \toprule
    \textbf{id} & \textbf{Model} & \textbf{Test w/o LM} &\textbf{Test with LM} \\
     & & \textbf{(WER)} & \textbf{(WER)} \\
    \midrule
    O1 & Mono.subword & 9.71 & 6.87 \\
    \hline
    M1 & \mbox{Whistle-small} + subword FT & \textbf{3.30} & \textbf{2.95} \\
    \hline
    M3 & \mbox{Wav2vec2-cv10} + subword FT & 3.76 & 3.06 \\
    \hline
    M4 & \mbox{Mul10-subword} + subword FT & 4.33 & 3.46 \\
    \bottomrule
  \end{tabular}
  
\end{table}

\subsection{Results for phoneme-based FT}

The Whistle model, during phoneme-based supervised pretraining, chooses to remove diacritics from phoneme annotations and to split diphthongs to maximize the sharing of pronunciation knowledge among different languages. 
Notably, Iu Mien script is a phonographic writing system closely related to pronunciation.
So when training the phoneme-based speech recognition model for the Iu Mien language, diacritics are retained, and diphthongs are not split when using IPA symbols to annotate word pronunciation. 
In this manner, we could describe the pronunciation differences between different words as accurately as possible.
For example, in Iu Mien, the pronunciation of the subword 'hn' is /\textipa{\r*n}/, and the pronunciation of the subword 'n' is /n/. If the diacritics were removed, the pronunciation of these two subwords could not be well distinguished.

Table \ref{tab:phoneme-result} shows the results for phoneme-based FT, starting from different pre-trained models.
O2 denotes the phoneme-based monolingual baseline, i.e., trained from scratch.
M2 denotes the result from fine-tuning of Whistle-small.
M5 denotes the result from fine-tuning of the Wav2Vec2-cv10 model.
% In the phoneme-based speech recognition experiments for the Iu Mien language, As shown in Table \ref{tab:phoneme-result}, we compared the Iu Mien speech recognition model fine-tuned based on the Whistle-small model with the Iu Mien speech recognition model trained from scratch and the model fine-tuned based on Wav2Vec2-cv10.
% To make the experimental results more statistically significant, we conducted three cross-tests for each experiment and took the average as the final experimental result.
Again, we conducted three independent cross-validation tests for each experiment and took the average as the final experimental result. 
The results in Table \ref{tab:subword-result} and Table \ref{tab:phoneme-result} use the same three splits, so they are directly comparable.

%\begin{table}[h]
%    \caption{ P-VALUES for comparison between different models.}
%    \label{tab:example}
%    \centering
%    \begin{tabular}{cccc}
%        \toprule
%       \textbf{Model-id} & \textbf{PER} & \textbf{WER w/o LM} & \textbf{WER with LM} \\
%        \midrule
%        M1 VS M3  & - & 0.26 & 0.51 \\
%        M1 VS M4  & - & 0.0002 & 0.15 \\
%        M3 VS M4 & - & 0.01 & 0.31 \\
%        M2 VS M5  & 0.07 & - & 0.23\\
%        M1 VS M5 & - & - & 0.60 \\
%        \bottomrule
%    \end{tabular}
%\end{table}

\subsection{Analysis}
By comparing O1 and M1, as well as O2 and M2, the results show that fine-tuning the pre-trained Whistle-small model on the Iu Mien language can significantly improve the performance, compared to training the model with only Iu Mien language data without using a pre-trained model.

When conducting subword based fine-tuning, the Whistle-small model (M1) performs the best, compared to both self-supervised pre-training (M3) and subword based pre-training (M4).
% shows significant advantages compared to the method based on self-supervised pre-training and then fine-tuning. 
Presumably, this is due to the clear goal of phoneme based pre-training, which directly learns to discriminate different sounds in the input audio. 

In phoneme-based fine-tuning experiments, using the Whistle model as the backbone model (M2) gave better results, compared to using the Wav2vec2-cv10 model as the backbone (M5). 
Additionally, it is found that phoneme-based FT of Whistle-small (M2) outperforms subword-based FT of Whistle-small (M1), which suggests that phoneme-based FT helps the phoneme-based backbone to work better and
the matching in pre-training and fine-tuning is beneficial.
% Although the output space during fine-tuning did not fully align with the output space during Whistle pre-training, the Whistle model still demonstrated data efficiency under low-resource conditions. 

\begin{table}[t]
  \centering
  \caption{PERs and WERs for phoneme-based fine-tuning \newline(averaged over three independent cross-validation runs)}
  \label{tab:phoneme-result}
  % \centering
  \begin{tabular}{ cp{2cm}cc }
    \toprule
    \textbf{id} & \textbf{Model} & \textbf{Test} &\textbf{Test} \\
     & & \textbf{(PER)} & \textbf{(WER)} \\
    \midrule 
    O2 & Mono.phoneme & 4.22 & 4.69 \\ 
    \hline
    % \cline{1-4}
    M2 & \mbox{Whistle-small} + phoneme FT & \textbf{2.41} & \textbf{2.71} \\
    \hline
    % \cline{1-4}
    M5 & \mbox{Wav2vec2-cv10} + phoneme FT & 2.53 & 2.76 \\
    \bottomrule
  \end{tabular}
\end{table}

\section{Conclusion and future work}

This paper primarily investigates and compares the effectiveness of different pre-training methods in fine-tuning Iu Mien language speech recognition models using a small amount of Iu Mien language data. By using the weakly-supervised phoneme-based multilingual pre-training model, Whistle, for fine-tuning, the speech recognition performance for the Iu Mien language can be effectively improved, achieving better results compared to other pre-training methods. Thanks to the common pronunciation among different languages, phoneme based pre-training with non-Chinese-Tibetan languages can also successfully improve the performance on the Chinese-Tibetan Iu Mien language. 
% The pre-training and fine-tuning method is found to solve the problem of insufficient training data for Iu Mien Language speech recognition.
Hopefully, this could reduce the data requirement for low-resourced speech recognition for a broader range of languages.

There are some interesting future work.
% First, annotated audio data for the Iu Mien language is relatively scarce. This paper only used about 9.7 hours of data for training and testing. Therefore, how to collect more data and how to perform data augmentation is a worthwhile fut
% the primary task at present. There have been some effort towards addressing this problem, such as using automated methods to collect audio and generate pseudo-labels \cite{yang2024gigaspeech2evolvinglargescale}. 
The Whistle model used in this paper was pre-trained using only ten non-tonal languages, without considering how to integrate tonal information from different languages into the multilingual pre-training. The Iu Mien language, however, has eight tones. There have been some effort towards addressing this problem \cite{li20k_interspeech,li2022autosegmental}.
Potentially, multilingual pre-training with tone modeling incorporated would benefit low-resourced speech recognition for tonal languages.

\bibliographystyle{IEEEtran}
\bibliography{main.bbl}

\begin{thebibliography}{10}
\providecommand{\url}[1]{#1}
\csname url@samestyle\endcsname
\providecommand{\newblock}{\relax}
\providecommand{\bibinfo}[2]{#2}
\providecommand{\BIBentrySTDinterwordspacing}{\spaceskip=0pt\relax}
\providecommand{\BIBentryALTinterwordstretchfactor}{4}
\providecommand{\BIBentryALTinterwordspacing}{\spaceskip=\fontdimen2\font plus
\BIBentryALTinterwordstretchfactor\fontdimen3\font minus \fontdimen4\font\relax}
\providecommand{\BIBforeignlanguage}[2]{{%
\expandafter\ifx\csname l@#1\endcsname\relax
\typeout{** WARNING: IEEEtran.bst: No hyphenation pattern has been}%
\typeout{** loaded for the language `#1'. Using the pattern for}%
\typeout{** the default language instead.}%
\else
\language=\csname l@#1\endcsname
\fi
#2}}
\providecommand{\BIBdecl}{\relax}
\BIBdecl

\bibitem{fromkin2007introduction}
V.~Fromkin, R.~Rodman, and N.~Hyams, \emph{An introduction to language: Eight edition}.\hskip 1em plus 0.5em minus 0.4em\relax Thomson Wadsworth, 2007.

\bibitem{yusuyin2024whistle}
S.~Yusuyin, T.~Ma, H.~Huang, W.~Zhao, and Z.~Ou, ``Whistle: Data-efficient multilingual and crosslingual speech recognition via weakly phonetic supervision,'' \emph{arXiv preprint arXiv:2406.02166}, 2024.

\bibitem{8682256}
H.~Xiang and Z.~Ou, ``\uppercase{CRF}-based single-stage acoustic modeling with \uppercase{CTC} topology,'' in \emph{International Conference on Acoustics, Speech, and Signal Processing (ICASSP)}, 2019.

\bibitem{zheng2021advancing}
H.~Zheng, W.~Peng, Z.~Ou, and J.~Zhang, ``Advancing \uppercase{CTC-CRF} based end-to-end speech recognition with wordpieces and conformers,'' \emph{arXiv preprint arXiv:2107.03007}, 2021.

\bibitem{soltau17_interspeech}
H.~Soltau, H.~Liao, and H.~Sak, ``{Neural Speech Recognizer: Acoustic-to-Word LSTM Model for Large Vocabulary Speech Recognition},'' in \emph{Proc. INTERSPEECH}, 2017.

\bibitem{10.5555/3618408.3619590}
A.~Radford, J.~W. Kim, T.~Xu, G.~Brockman, C.~McLeavey, and I.~Sutskever, ``Robust speech recognition via large-scale weak supervision,'' in \emph{International Conference on Machine Learning (ICML)}, 2023.

\bibitem{Babu2021XLSRSC}
A.~Babu, C.~Wang, A.~Tjandra, K.~Lakhotia, Q.~Xu, N.~Goyal, K.~Singh, P.~von Platen, Y.~Saraf, J.~M. Pino, A.~Baevski, A.~Conneau, and M.~Auli, ``\uppercase{XLS-R}: Self-supervised cross-lingual speech representation learning at scale,'' in \emph{Proc. INTERSPEECH}, 2021.

\bibitem{baevski2020wav2vec}
A.~Baevski, Y.~Zhou, A.~Mohamed, and M.~Auli, ``wav2vec 2.0: A framework for self-supervised learning of speech representations,'' \emph{Advances in neural information processing systems}, vol.~33, pp. 12\,449--12\,460, 2020.

\bibitem{kenton2019bert}
J.~D. M.-W.~C. Kenton and L.~K. Toutanova, ``\uppercase{BERT}: Pre-training of deep bidirectional transformers for language understanding,'' in \emph{Conference of the North American Chapter of the Association for Computational Linguistics: Human Language Technologies (NAACL)}, 2019.

\bibitem{graves2006connectionist}
A.~Graves, S.~Fern{\'a}ndez, F.~Gomez, and J.~Schmidhuber, ``Connectionist temporal classification: labelling unsegmented sequence data with recurrent neural networks,'' in \emph{Proceedings of the 23rd international conference on Machine learning (ICML)}, 2006.

\bibitem{ardila2020common}
R.~Ardila, M.~Branson, K.~Davis, M.~Kohler, J.~Meyer, M.~Henretty, R.~Morais, L.~Saunders, F.~Tyers, and G.~Weber, ``Common voice: A massively-multilingual speech corpus,'' in \emph{Proceedings of the Twelfth Language Resources and Evaluation Conference (LREC)}, 2020.

\bibitem{An2020CATAC}
K.~An, H.~Xiang, and Z.~Ou, ``\uppercase{CAT}: A \uppercase{CTC-CRF} based \uppercase{ASR} toolkit bridging the hybrid and the end-to-end approaches towards data efficiency and low latency,'' in \emph{Proc. INTERSPEECH}, 2020.

\bibitem{kudo2018sentencepiece}
T.~Kudo and J.~Richardson, ``Sentencepiece: A simple and language independent subword tokenizer and detokenizer for neural text processing,'' in \emph{Proceedings of the 2018 Conference on Empirical Methods in Natural Language Processing: System Demonstrations (EMNLP)}, 2018.

\bibitem{heafield2011kenlm}
K.~Heafield, ``Kenlm: Faster and smaller language model queries,'' in \emph{Proceedings of the sixth workshop on statistical machine translation (WMT)}, 2011.

\bibitem{mohri2008speech}
M.~Mohri, F.~Pereira, and M.~Riley, ``Speech recognition with weighted finite-state transducers,'' \emph{Springer Handbook of Speech Processing}, pp. 559--584, 2008.

\bibitem{li20k_interspeech}
J.~Li and M.~Hasegawa-Johnson, ``{Autosegmental Neural Nets: Should Phones and Tones be Synchronous or Asynchronous?}'' in \emph{Proc. INTERSPEECH}, 2020.

\bibitem{li2022autosegmental}
------, ``Autosegmental neural nets 2.0: An extensive study of training synchronous and asynchronous phones and tones for under-resourced tonal languages,'' \emph{IEEE/ACM Transactions on Audio, Speech, and Language Processing}, vol.~30, pp. 1918--1926, 2022.

\end{thebibliography}

% \begin{thebibliography}{9}
% \bibitem[1]{Davis80-COP}
%   S.\ B.\ Davis and P.\ Mermelstein,
%   ``Comparison of parametric representation for monosyllabic word recognition in continuously spoken sentences,''
%   \textit{IEEE Transactions on Acoustics, Speech and Signal Processing}, vol.~28, no.~4, pp.~357--366, 1980.
% \bibitem[2]{Rabiner89-ATO}
%   L.\ R.\ Rabiner,
%   ``A tutorial on hidden Markov models and selected applications in speech recognition,''
%   \textit{Proceedings of the IEEE}, vol.~77, no.~2, pp.~257-286, 1989.
% \bibitem[3]{Hastie09-TEO}
%   T.\ Hastie, R.\ Tibshirani, and J.\ Friedman,
%   \textit{The Elements of Statistical Learning -- Data Mining, Inference, and Prediction}.
%   New York: Springer, 2009.
% \bibitem[4]{YourName17-XXX}
%   F.\ Lastname1, F.\ Lastname2, and F.\ Lastname3,
%   ``Title of your ISCSLP 2024 publication,''
%   in \textit{ISCSLP 2024 -- 23\textsuperscript{rd} Annual Conference of the International Speech Communication Association, September 18-22, Incheon, Korea, Proceedings, Proceedings}, 2024, pp.~100--104.
% \end{thebibliography}

\end{CJK*}

\end{document}